\documentclass[aps,prl,twocolumn,showpacs]{revtex4}
\begin{document}
\title{Comment on ``Off-diagonal Long-range Order in Bose Liquids:
Irrotational Flow and Quantization of Circulation''
 }
\author{Yu Shi}
\email{ys219@phy.cam.ac.uk}
\affiliation{Cavendish Laboratory,
University of Cambridge, Cambridge CB3 0HE, United Kingdom}
\begin{abstract}
In the context of an application to superfluidity, it is
elaborated how to do quantum mechanics of a system with a
rotational velocity. Especially, in both the laboratory frame and
the non-inertial co-rotating frame, the canonical momentum, which
corresponds to the quantum mechanical momentum operator, contains
a part due to the rotational velocity.
\end{abstract}

\maketitle

A recent Letter~\cite{su} claimed to have derived irrotational
superflow or circulation quantization of a superfluid  merely from
the existence of off-diagonal long-range order (ODLRO). The
derivation  was based on an incorrect Hamiltonian.

In the beginning of that discussion, it is  written: ``Let us
consider a bucket of Bose liquid composed of $N$ homogeneous
interacting spinless particles, rotated with a constant angular
velocity $\mathbf{\Omega}$''~\cite{su}.  This clearly indicates
that $\mathbf{v}_s = \mathbf{\Omega}\times\mathbf{r}$ is the fixed
rotation velocity of the bucket, not the  superfluid velocity.
$\mathbf{v}_s$ and $\mathbf{\Omega}$ are both physical quantities,
and one has no ``gauge freedom'' to add an additional term
$\nabla\theta(\mathbf{r})$ in $\mathbf{v}_s$, as appearing in
Eq.~(2) of \cite{su}. With this understanding, the conclusion that
$\mathbf{\Omega}$ vanishes is, of course,  absurd. In fact, for a
superfluid in a rotating bucket, it is  the Hamiltonian in the
co-rotating frame that determines the equilibrium statistical
mechanics~\cite{landau,pines}. Instead of Eq.~(1) of \cite{su},
the relevant Hamiltonian  is $\sum_{j}
[\frac{(\mathbf{p}_j-m\mathbf{\Omega}\times\mathbf{r}_j)^2}{2m}
-\frac{1}{2}m(\mathbf{\Omega}\times\mathbf{r}_j)^2]
+\frac{1}{2}\sum_{j\neq l}V(r_{jl})$,  where
$\mathbf{p}_j=m(\mathbf{v}_j+\mathbf{\Omega}\times\mathbf{r}_j)$
is the canonical momentum and is then replaced as
$-i\hbar\nabla_j$,
 $\mathbf{v}_j$, $\mathbf{r}_j$ and $\mathbf{p}_j$
are relative to  the reference frame co-rotating with
$\mathbf{\Omega}$~\cite{landau,pines,ll,lq}. This Hamiltonian
obviously lacks the invariance under an
$\mathbf{\Omega}\times\mathbf{r}$-dependent re-gauged translation
underlying the argument employed later  in \cite{su}.

In their correspondence, the authors have claimed that they were
not considering   superfluid in a rotating bucket, instead,
$\mathbf{v}_s= \mathbf{\Omega}\times\mathbf{r}+ \nabla\theta$ was
a ``drift velocity'' assumed to be a sum of
 an irrotational part $\nabla\theta$ and a rotational part
$\mathbf{\Omega}\times\mathbf{r}$. Under this interpretation, even
though the serious question about the physical meaning of this
``drift velocity'' is side-stepped, the Hamiltonian in  \cite{su}
is still incorrect. In fact, whenever a certain  ``drift
velocity'' contains a rotational part
$\mathbf{\Omega}\times\mathbf{r}$, the canonical momentum relative
to the non-inertial frame co-moving with this  ``drift velocity''
is
$\mathbf{p}_j=m(\mathbf{v}_j+\mathbf{\Omega}\times\mathbf{r}_j)$,
and thus in the laboratory frame, the  kinetic part of the
Hamiltonian
 $(\mathbf{p}_j+m\nabla\theta)^2/2m$~\cite{ll},
in contrast with Eq.~(1) of \cite{su}.
It is very elementary knowledge that it is canonical
momentum that corresponds to momentum operator in quantum
mechanics~\cite{ll,lq}.
Moreover,  in \cite{su}, it is very mysterious  why the assumed
``drift velocity'' suddenly became the superfluid velocity after the
the rotational part was ``shown'' to vanish.

If the authors  wish  to adopt a {\it reductio ad absurdam} by
showing that ODLRO is incompatible with the assumption that the
superflow has a non-vanishing curl, i.e. the superfluid velocity
is rotational, then the only valid interpretation is that
$\mathbf{v}_s$ represents  the superfluid velocity, first {\em
presumed} to has a curl $2\mathbf{\Omega}$; the purpose is then to
show $\mathbf{\Omega}$ has to vanish under ODLRO. If we use a
subscript $0$ to represent the laboratory frame, then the
assumption is $\frac{1}{2}\nabla_0\times \mathbf{v}_s =
\mathbf{\Omega}$. With this interpretation, and  if
$\mathbf{\Omega}$ is assumed to be position-independent, one can
write $\mathbf{v}_s = \mathbf{\Omega}\times\mathbf{r}_0 +
\nabla_0\theta(\mathbf{\Omega},\mathbf{r}_0)$. In terms of the 
position vector $\mathbf{r}$ relative to the condensate, which
moves with  $\mathbf{v}_s$, the expression of $\mathbf{v}_s$
becomes very complicated.   In general, Eq.~(2) in \cite{su}   is
valid only if $\mathbf{r}$ and thus $\mathbf{p}_j$ in \cite{su}
are relative to the rest frame. But this understanding cannot be
consistent with the Hamiltonian in \cite{su}. In fact, the
Hamiltonian in the rest frame, in our  notations,  is simply
$\sum_{j} \frac{{{\mathbf{p}_j}_0}^2}{2m} +\frac{1}{2}\sum_{j\neq
l}V(r_{jl})$, where ${{\mathbf{p}_j}_0}=-i\hbar{\nabla_j}_0$,
instead of Eq.~(1) of \cite{su}. If one subsequently obtains (1)
by a ``gauge'' transformation, then $m\mathbf{v}_s/\hbar$ is only
the gradient of the phase factor, and $\mathbf{v}_s$ cannot be the
superfluid velocity if it is presumed to be rotational as in (2)
(cf. the preceding paragraph). On the other hand, it appears that
$\mathbf{p}_j$ in \cite{su} are relative to the condensate and
that the authors wish to give the Hamiltonian in the rest frame
expressed  in terms of the momenta relative to the condensate.
With this interpretation,  however, both Eq.~(2) in \cite{su}  and
the expression $\mathbf{\Omega}=\frac{1}{2}\nabla \times
\mathbf{v}_s$ below the equation are not valid generically under
the presumption that the curl of the superfluid velocity is
$\mathbf{\Omega}$  in the laboratory frame. Therefore  the
Hamiltonian in \cite{su} is still incorrect.

It appears that the authors
do not know that ``irrotational superflow '' just means
that the curl of the superfluid velocity vanishes, in the laboratory
frame, and that the purpose  they had set for their Letter
was just to show that with ODLRO,   the superfluid
velocity is of the form $\nabla\theta$, i.e. its curl vanishes.
Moreover, they ignore the well-known fundamental difference between rotation
and translation~\cite{landau,pines,ll,lq}.  Note that
in Sec.~$5.5$ (pp. $75-76$) of Ref.~\cite{pines},
it is a {\em uniform}  (as italicized  there)
translation of a superfluid  that is considered.

I thank  Nigel Cooper  for useful discussions.


\begin{thebibliography}{9}
\bibitem{su} G. Su and M. Suzuki, Phys. Rev. Lett. {\bf 86}, 2708 (2001).
\bibitem{landau}  L. D. Landau and E. M. Lifshitz, {\em Statistical Mechanics}
(Pergamon Press, Oxford, 1976).
\bibitem{pines} P. Nozi\`{e}res and D. Pines
 {\em The Theory of Quantum Liquids, Vol.~2}
(Addison-Wesley, Redwood City, 1990).
\bibitem{ll}  L. D. Landau and E. M. Lifshitz, {\em Mechanics}
(Pergamon Press, Oxford, 1976), Sec.~39.
\bibitem{lq} L. D. Landau and E. M. Lifshitz, {\em Quantum  Mechanics}
(Pergamon Press, Oxford, 1976).

\end{thebibliography}
\end{document}